\documentclass[twocolumn,showpacs,amsmath,amssymb,nofootinbib]{revtex4-1}
\usepackage{dcolumn}
\usepackage{bm}
\usepackage{graphicx}
\usepackage{color}

\usepackage{amsthm}

\newcommand{\be}{\begin{equation}}
\newcommand{\ee}{\end{equation}}
\newcommand{\ba}{\begin{eqnarray}}
\newcommand{\ea}{\end{eqnarray}}
\newcommand{\non}{\nonumber}

\newtheorem*{theorem}{Theorem}

\begin{document}

\title{Local Invariants Vanishing on Stationary Horizons:
\\A Diagnostic for Locating Black Holes}
\author{Don N. Page}
\email{profdonpage@gmail.com}
\author{Andrey A. Shoom}
\email{ashoom@ualberta.ca}
\affiliation{Theoretical Physics Institute, University of Alberta,
Edmonton, Alberta T6G 2E1, Canada}

\date{2015 March 19}

\begin{abstract}
Inspired by the example of Abdelqader and Lake for the Kerr metric, we
construct local scalar polynomial curvature invariants that vanish on
the horizon of any stationary black hole:  the squared norms of the
wedge products of $n$ linearly independent gradients of scalar polynomial
curvature invariants, where $n$ is the local cohomogeneity of the spacetime.
\end{abstract}

\pacs{04.70.Bw, 04.50.Gh, 04.20.Cv, 02.40.Hw \hfill Alberta-Thy-1-15, arXiv:1501.03510}

\maketitle

How may one locate a black hole?  This question may be quite challenging even for numerical computation, for the location of a black hole horizon is a delicate issue.  Here we propose a method which generically gives the precise location of any stationary horizon and should give the approximate location of a nearly stationary horizon.  Hence it should be useful for numerical computations in general relativity.

For a general black hole spacetime, the location of its surface (the event horizon, which is the boundary of the region from which causal curves can go to asymptotic future null infinity) depends on the future evolution of the spacetime and is not determined locally.  However, for black holes that settle down to become stationary, one might ask whether one can find a local invariant that is generically nonzero off the horizon but vanishes on the horizon.  In particular, one might look for a scalar polynomial curvature invariant \cite{Stephani:2003tm,Lake:2003qe,Coley:2009eb,Hervik:2010rg,Coley:2010ze,Lake:2012ew,Abdelqader:2013bma}, which is a scalar obtained by complete contraction of all the indices of a polynomial in the Riemann curvature tensor $R_{\alpha\beta\gamma\delta}$ and its covariant derivatives.  For example, Anders Karlhede, Ulf Lindstr\"{o}m and Jan Aman \cite{Karlhede:1982fj} showed that $R^{\alpha\beta\gamma\delta;\epsilon}R_{\alpha\beta\gamma\delta;\epsilon}$ crosses zero and switches sign as one crosses the horizon of the Schwarzschild metric, and one can easily show that this is also true for any smooth static spherically symmetric horizon.  An invariant that vanishes on more general stationary horizons could be useful in numerical relativity for finding the approximate location of the horizon once the spacetime has settled down to become approximately stationary.  

Majd Abdelqader and Kayll Lake \cite{Abdelqader:2014vaa} have recently found a local scalar polynomial curvature invariant that vanishes on the horizon of the Kerr black hole.  After casting this invariant into a simpler form that is proportional to the squared norm of the wedge product of two curvature-invariant gradients, we realized that the procedure generalizes to give a way to locate any nonsingular stationary horizon in terms of local curvature invariants.  Essentially, if one constructs as many gradients of independent curvature invariants as the local cohomogeneity of a generic stationary spacetime, at a generic point in the spacetime outside an horizon these gradients will be linearly independent and spacelike, but on the horizon, a linear combination will become null.  This implies that the squared norm of the wedge product of the gradients will vanish on a stationary horizon.

Our procedure does not assume the Einstein equations, but only a smooth stationary horizon (a null hypersurface generated by a Killing vector field that is timelike outside the horizon but null on the horizon).  Therefore, in principle the invariants to be used in the procedure could include those from the Ricci tensor $R_{\alpha\beta}$, such as the Ricci scalar $R = R^{\alpha}_{\alpha}$ or $R^{\alpha\beta}R_{\alpha\beta}$.  However, because the main interest may be in Ricci-flat ($R_{\alpha\beta} = 0$) spacetimes that solve the vacuum Einstein equations, in the examples below we shall consider only invariants obtained from the Riemann curvature tensor $R_{\alpha\beta\gamma\delta}$ (or from the Weyl tensor $C_{\alpha\beta\gamma\delta}$, which is the trace-free part of the Riemann tensor) that need not vanish even when $R_{\alpha\beta} = 0$.

In the original (30 Dec.\ 2014) version of their paper, Abdelqader and Lake \cite{Abdelqader:2014vaa} gave the following six curvature invariants for the Kerr metric, which we here copy directly from that paper:
\begin{equation}
 I_1 \equiv C_{\alpha \beta \gamma \delta}\;C^{\alpha \beta \gamma \delta} \label{defi1} \;,
\end{equation}
\begin{equation}
 I_2 \equiv {C^{*}}_{\alpha \beta \gamma \delta}\;C^{\alpha \beta \gamma \delta}\label{defi2} \;,
\end{equation}
\begin{equation}
 I_3 \equiv \nabla_{\mu} C_{\alpha \beta \gamma \delta} \;\nabla^{\mu} C^{\alpha \beta \gamma \delta}\label{defi3}\;,
 \end{equation}
\begin{equation}
 I_4 \equiv \nabla_{\mu} C_{\alpha \beta \gamma \delta} \;\nabla^{\mu} {C^{*}}^{\; \alpha \beta \gamma \delta}\label{defi4}\;,
\end{equation}
\begin{equation}
I_5 \equiv k_{\mu} k^{\mu}\label{defi5}\;,
\end{equation}
and
\begin{equation}
 I_6 \equiv {l}_{\mu} {l}^{\mu}\label{defi6}\;,
\end{equation}
where $ {C}_{\alpha \beta \gamma \delta} $ is the Weyl tensor, $ {C^{*}}_{\alpha \beta \gamma \delta} $ its dual, $ k_{\mu} \equiv - \nabla_{\mu}\, I_1\,$, and $ {l}_{\mu} \equiv - \nabla_{\mu}\, I_2 \;$.
Then they showed that in the Kerr metric the dimensionless scalar nonpolynomial curvature invariant (with scalar polynomial curvature invariant numerator)
\begin{equation}
Q_2 \equiv \frac{\left(I_5 +I_6 \right)^2-(12/5)^2 \left( {I_1}^2 +{I_2}^2 \right) \left( {I_3}^2 +{I_4}^2 \right) }{108\left( {I_1}^2 +{I_2}^2 \right)^{5/2}}  \label{q2}
\end{equation}
vanishes on the black hole horizon.

We found that the syzygy (in this context a functional relationship between invariants for a particular spacetime geometry) that Abdelqader and Lake \cite{Abdelqader:2014vaa} discovered for the Kerr metric, that
\begin{equation}\label{syzygy1}
I_6-I_5+ \frac{12}{5} \left(I_1\,I_3-I_2\,I_4 \right)=0 \;,
\end{equation}
may be expressed as the real part of the complex syzygy
\begin{equation}\label{syzygy1'}
\nabla_{\mu}\left(I_1+iI_2\right)\nabla^{\mu}\left(I_1+iI_2\right) =
\frac{12}{5} \left(I_1+iI_2\right) \left(I_3+iI_4\right) \;
\end{equation}
whose imaginary part gives a syzygy for a seventh invariant,
\begin{equation}
 I_7 \equiv {k}_{\mu} {l}^{\mu} = \frac{6}{5} \left(I_1\,I_4+I_2\,I_3 \right)\label{defi7}\;.
\end{equation}
Using both the real and imaginary parts of this complex syzygy, one may readily see that
\begin{equation}\label{Q_2}
27 \left({I_1}^2+{I_2}^2 \right)^{5/2} Q_2 = 
(k_{\mu} k^{\mu})(l_{\nu} l^{\nu}) - (k_{\mu} l^{\mu})(l_{\nu} k^{\nu})\;,
\end{equation}
which is the squared norm of the wedge product $dI_1\wedge dI_2$ of the gradients of the Kretschmann invariant $I_1$ and of the Chern-Pontryagin invariant $I_2$ (for a Ricci-flat spacetime; otherwise in four dimensions the Kretschmann invariant is $R_{\alpha \beta \gamma \delta}R^{\alpha \beta \gamma \delta} = I_1 + 2 R_{\alpha \beta} R^{\alpha \beta} - (1/3) R^2$, though the Chern-Pontryagin invariant ${R^{*}}_{\alpha \beta \gamma \delta}\;R^{\alpha \beta \gamma \delta} = I_2$ has no correction from the Ricci tensor).

Although it is incidental to the main point of our paper, we also found another syzygy from a minor extension of the work of Abdelqader and Lake \cite{Abdelqader:2014vaa},
\begin{equation}\label{syzygy2}
\Im{\left[\left(I_1+iI_2\right)^4\left(I_3-iI_4\right)^3\right]} = 0 \;,
\end{equation}
where $\Im$ denotes the imaginary part of what follows.  (The real part is equal to the absolute value in this case and does not vanish to give yet another syzygy.)  With these three syzygies for the seven real scalar polynomial curvature invariants, $I_1, \ldots , I_7$, one is left with four independent invariants, precisely the right number to determine generically (up to eight sign choices) the values of the two Kerr parameters $M$ and $a$ and of the two nonignorable coordinates $r$ and $\theta$.  (None of these three remain syzygies when one adds a cosmological constant, so the syzygies for that case remain to be found.)

Eq.\ (\ref{Q_2}) for the Abdelqader-Lake invariant $Q_2$ that they found vanishes on the horizon of the Kerr metric inspired the realization that the squared norm of the wedge product of $n$ gradients of independent local smooth curvature invariants would vanish on the horizon of any stationary black hole, where $n$ is the local cohomogeneity of the spacetime (the codimension of the maximal dimensional orbits of the isometry group of the local metric, ignoring the breaking of any of these local isometries by global considerations, such as the way that the identifications of flat space to make a torus break the local rotational part of the isometry group), so that at generic points outside an horizon of a generic metric of local cohomogeneity $n$, the squared norm of the wedge product would be positive.  (We use the metric signature $-++\cdots$.)  Let us now prove that this squared norm vanishes on a stationary horizon.

Let $D$ be the dimensionality of the spacetime, and let $m$ be the maximal dimension of the orbits of the local isometry group.  Then $n = D-m$ is the local cohomogeneity of the spacetime.  For a  stationary spacetime of local cohomogeneity $n$ with an event horizon, let $\xi^{\mu}$ denote the Killing vector field that on the horizon is its null generator, which is orthogonal to the null horizon hypersurface.  In a neighborhood outside the horizon, $\xi^{\mu}$ will be timelike, with $\xi^{\mu}\xi_{\mu} < 0$ in our choice of signature.  Let $\{S^{(i)}\}$, for $i$ ranging from 1 to $n$, denote a set of $n$ functionally independent nonconstant scalar polynomial curvature invariants (total scalar contractions over all indices of polynomials of curvature tensors and of their covariant derivatives).  For example, some or all of the ${S^{(i)}}$ could be chosen from the set ${I_1\ldots I_7}$ given above (though remembering that $I_2$, $I_4$, $I_6$, and $I_7$ above are only defined for $D=4$), but one could also choose invariants from total scalar contractions of other polynomials in the Riemann curvature tensor and its covariant derivatives.

Then let $\{dS^{(i)}\}$ be the set of exterior derivatives (gradients) of the $n$ scalar polynomial curvature invariants, with components $S^{(i)}_{;\mu} = \nabla_{\mu}S^{(i)}$.  Since the curvature invariants are invariant under translations by the local isometry group, their gradients all lie with the $n$-dimensional local cohomogeneity part of the cotangent space at each point \cite{Lake:2012ew}.  The wedge product of $n$ gradients, the $n$-form $W = dS^{(1)}\wedge dS^{(2)}\wedge\cdots\wedge dS^{(n)}$, will be proportional to the volume form in this $n$-dimensional part of the cotangent space at each point of spacetime, with a proportionality depending upon the spacetime point.  At generic points in a generic spacetime, the proportionality will be nonzero where the gradients of the $n$ scalar invariants are linearly independent, but there can be a set of points (generically hypersurfaces) where the $n$ gradients are not linearly independent, so that the wedge product vanishes there and the proportionality is zero.

The Hodge dual $*W$ is then an $m$-form that at each point of the spacetime lies entirely within the $m$-dimensional part of the cotangent space that is generated by the local isometries.  It will be proportional to the $m$-dimensional volume element of that part of the cotangent space, though the location-dependent proportionality will be zero at the same set of spacetime points at which the wedge product $W$ vanishes.

On a stationary horizon, which is a null hypersurface generated by the Killing vector $\xi^{\mu}$ that is timelike outside the horizon but null on the horizon (and hence both parallel and orthogonal to the horizon), the $m$-dimensional part of the cotangent space will include the null one-form, with one-form components $\xi_{\mu} = g_{\mu\nu}\xi^{\nu}$, which is metrically equivalent to the Killing vector $\xi^{\mu}$ that is null and hypersurface-orthogonal at the horizon, as well as other spacelike one-forms metrically equivalent to other Killing vectors at the horizon that are all spacelike if $m>1$.  Therefore, this $m$-dimensional part of the cotangent space will be null on the stationary horizon, and hence $*W$ will also be null, having zero squared norm.  Since the squared norm of $W$, appropriately defined, is the same as that of its Hodge dual $*W$, the squared norm of the wedge product $W$ of $n$ scalar polynomial curvature invariants will be zero on a stationary horizon.

There is of course in this argument the implicit assumption that the spacetime is sufficiently regular at the event horizon that the gradients of the $n$ scalar polynomial curvature invariants are well-defined at the horizon, so that their wedge product $W$ is well-defined and finite.

We summarize this result by the following theorem:

\begin{theorem}
For a spacetime of local cohomogeneity $n$ that contains a stationary horizon (a null hypersurface that is orthogonal to a Killing vector field that is null there and hence lies within the hypersurface and is its null generator) and which has $n$ scalar polynomial curvature invariants ${S^{(i)}}$ whose gradients are well-defined there, the $n$-form wedge product $W = dS^{(1)}\wedge dS^{(2)}\wedge\cdots\wedge dS^{(n)}$ has zero squared norm on the horizon,
\ba
 \|W\|^2&\equiv&\frac{1}{n!}\delta^{\alpha_1\ldots\alpha_n}_{\beta_1\ldots\beta_n}
 g^{\beta_1\gamma_1}\cdots g^{\beta_n\gamma_n}\non\\
 &\times&S^{(1)}_{;\alpha_1}\cdots S^{(n)}_{;\alpha_n}
 S^{(1)}_{;\gamma_1}\cdots S^{(n)}_{;\gamma_n} = 0
 \label{norm}\;.
\ea
where the permutation tensor $\delta^{\alpha_1\ldots\alpha_n}_{\beta_1\ldots\beta_n}$ is $+1$ if $\alpha_1\ldots\alpha_n$ is an even permutation of $\beta_1\ldots\beta_n$, is $-1$ if $\alpha_1\ldots\alpha_n$ is an odd permutation of $\beta_1\ldots\beta_n$, and is zero otherwise (including all cases of repeated indices upstairs or downstairs).
\end{theorem}

Then $\|W\|^2$ is itself a scalar polynomial curvature invariant that vanishes on any stationary horizon smooth enough for $W$ to be well-defined there.  Of course, for this to be useful for locating a stationary horizon, the $n$ scalar polynomial curvature invariants ${S^{(i)}}$ should be chosen to be functionally independent so that $\|W\|^2$ is positive at generic points in the spacetime outside the horizon.  This will not always be possible, such as for the cosmological horizon of de Sitter spacetime, which is totally homogeneous and hence has cohomogeneity $n=0$, though of course for this spacetime there are cosmological horizons running through every point.

This raises the following question:  For a spacetime of local cohomogeneity $n$ with a Killing horizon that is a locally unique hypersurface\footnote{By locally unique, we mean that any neighborhood of the horizon includes points not on a Killing horizon, unlike the case for de Sitter or certain other examples, such as spacetimes having a covariantly constant null vector field, that have Killing horizons through all points.}, are there always enough functionally independent scalar polynomial curvature invariants so that one can choose $n$ of them in such a way that the squared norm of the wedge product of their gradients is generically nonvanishing away from the horizon?  If the answer is `no,' then although our theorem above would remain true, the squared norm of the wedge product of the gradients of any $n$ scalar polynomial curvature invariants would vanish in a whole neighborhood of a stationary horizon, so it would not be useful for locating the horizon.  Based upon the work of Alan Coley, Sigbj{\o}rn Hervik, and Nicos Pelavas \cite{Coley:2009eb, Hervik:2010rg, Coley:2010ze}, one might conjecture that the answer to the question is `yes,' but so far we do not have a rigorous proof.

Let us now consider various examples in four-dimensional spacetimes of the fact that the squared norm of the wedge product of $n$ gradients of curvature invariants vanishes at the horizon of a stationary black hole.

First, consider a spherically symmetric static black hole, so the codimension is $n=1$ (the radial direction, say with coordinate $r$).  Then any smooth curvature invariant, such as the Kretschmann invariant $R_{\alpha \beta \gamma \delta}R^{\alpha \beta \gamma \delta}$, which is the same as $I_1$ for a Ricci-flat spacetime, will have a gradient in the radial direction that becomes null on the horizon (e.g., as seen in a frame parallely propagated by an infalling geodesic), so the square of the norm of the gradient of the Kretschmann invariant (with only one gradient in the wedge product in this $n=1$ case) will be zero on any spherically symmetric static black hole horizon.

Second, consider a stationary axisymmetric black hole in four dimensions, which has two commuting Killing vector fields and cohomogeneity $n=2$, such as the Kerr metric.  Then we need the wedge product of two gradients for its squared norm to be positive at generic locations outside the black hole but vanishing on the horizon.  For a rotating black hole such as Kerr, one can take the squared norm of the wedge product $dI_1\wedge dI_2$ as Abdelqader and Lake \cite{Abdelqader:2014vaa} effectively did (though originally without realizing explicitly that their invariant $Q_2$ is proportional to the squared norm of this wedge product before we discovered and pointed out this fact).  For a static axisymmetric black hole, the invariants $I_2$, $I_4$, $I_{6}$, and $I_7$ vanish, so one could instead take the squared norm of $dI_1\wedge dI_3$, of $dI_1\wedge dI_5$, of $dI_3\wedge dI_5$, or of any other pair of independent nonvanishing scalar polynomial curvature invariants, such as the trace of higher powers of the curvature tensor like $R^{\alpha\beta}_{\ \ \ \gamma\delta}R^{\gamma\delta}_{\ \ \ \epsilon\zeta}R^{\epsilon\zeta}_{\ \ \ \alpha\beta}$.  

In these cases with more than one Killing vector that all commute (so that the dimensionality $m$ of the isometry group is the same as the number of Killing vectors), the squared norm of the wedge product of $n$ independent scalar polynomial curvature invariants will also vanish at the fixed points of the Killing vectors other than the one that becomes null on the horizon, such as on the axes of axisymmetric spacetimes.  In such cases we do not know how to construct scalar polynomial curvature invariants that vanish only on the horizon, so that will be left as a challenge for the future, as well as the challenge as to whether for specific spacetimes such as Kerr, there is a scalar polynomial curvature invariant without any covariant derivatives of the Riemann curvature tensor that vanishes on the horizon or that vanishes only on the horizon.

Third, consider a distorted static black hole in four dimensions that has no spatial Killing vector fields on and outside the horizon, so the local cohomogeneity is $n=3$.  Then we need the squared norm of the wedge product of three gradients of scalar polynomial curvature invariants, such as $dI_1\wedge dI_3\wedge dI_5$ or wedge products using the gradients of the traces of higher powers of the curvature tensor and/or of its covariant derivatives.

There are of course analogous examples in spacetimes of higher dimension $D > 4$.  If they are static and spherically symmetric, then for any dimension $D \geq 4$, the local cohomogeneity is $n=1$, and the gradient of any scalar polynomial curvature invariant (such as the Kretschmann invariant) that has a nonvanishing gradient just outside the horizon will have its squared norm becoming zero on the horizon.

The general Kerr-NUT-(A)dS metric in spacetime dimension $D$ \cite{Chen:2006xh} has local cohomogeneity $n = \lfloor D/2 \rfloor$, the greatest integer not greater than $D/2$.  That is, the $D=4$ case has two commuting Killing vector fields and local cohomogeneity $n=2$ (as the special case of the Kerr metric does with both the NUT parameter and the cosmological constant zero).  Analogously, the $D=5$ case has three commuting Killing vector fields (one becoming null on the horizon and two axial ones) and also $n = \lfloor 5/2 \rfloor = 2$.  Then $D=6$ and $D=7$ have $n=3$, $D=8$ and $D=9$ have $n=4$, $D=10$ and $D=11$ have $n=5$, etc.  In each case the squared norm of the wedge product of the gradients of $n$ scalar polynomial curvature invariants will vanish on the horizons (as well as on the axes surfaces, the fixed points of the axial Killing vector fields).

One might conjecture that one could take the traces of the 2nd through the $(n+1)$th powers of the Riemann curvature tensor as one choice of $n$ invariants.  We have looked at sample values of the 3 parameters (mass $M$, rotation parameter $a$, and NUT parameter $L$ at fixed cosmological constant) of the Kerr-NUT-(A)dS metric in spacetime dimension $D=4$ and have found that indeed the square of the norm of the wedge product of the traces of the 2nd and 3rd powers of the Riemann curvature tensor (the Kretschmann invariant $R_{\alpha \beta \gamma \delta}R^{\alpha \beta \gamma \delta}$ and $R^{\alpha\beta}_{\ \ \ \gamma\delta}R^{\gamma\delta}_{\ \ \ \epsilon\zeta}R^{\epsilon\zeta}_{\ \ \ \alpha\beta}$) is generically nonzero away from the horizons and axes, but there are curves in the $(r,\theta)$ plane (hypersurfaces of the full spacetime that are neither horizons nor axes) where the square of the norm of the wedge product vanishes.  

In particular, when the NUT parameter $L$ is zero, the Kerr-(A)dS spacetime has a reflection symmetry about the equatorial plane hypersurface, if one ignores the change in sign of the volume element given by the totally antisymmetric Levi-Civita tensor.  In this case, gradients of scalar polynomial curvature invariants that do not involve odd powers of the Levi-Civita tensor (e.g., not counting $I_2$, $I_4$, and $I_7$ in $D=4$) will lie in the $(n-1)$-dimensional local cohomogeneity part of the hypersurface of reflection symmetry, so that the wedge product of any $n$ such gradients will vanish there.  When the NUT parameter is nonzero, there is no reflection symmetry, but generically there remain distorted hypersurfaces where the wedge product vanishes.

We found that there are generically other hypersurfaces away from the horizons and axes on which the wedge product of $n = 2$ gradients vanishes.  However,  we found that if we took the three scalar polynomial curvature invariants $S^{(1)} = R_{\alpha \beta \gamma \delta}R^{\alpha \beta \gamma \delta}$, $S^{(2)} = R^{\alpha\beta}_{\ \ \ \gamma\delta}R^{\gamma\delta}_{\ \ \ \epsilon\zeta}R^{\epsilon\zeta}_{\ \ \ \alpha\beta}$, and $S^{(3)} = R_{\alpha \beta \gamma \delta ; \epsilon}R^{\alpha \beta \gamma \delta ; \epsilon}$, then we have three pairs of wedge products, $W_{12} = dS^{(1)}\wedge dS^{(2)}$, $W_{23} = dS^{(2)}\wedge dS^{(3)}$, and $W_{31} = dS^{(3)}\wedge dS^{(1)}$, that each vanish on certain curves in the $(r,\theta)$ plane and with each pair concurrently vanishing at certain intersection points in the $(r,\theta)$ plane, but with no triple intersections where all three wedge products concurrently vanish for the sample values of $M$, $a$, and $L$ that we chose.  Therefore, it appears that if one takes the sum of the three squared norms of the wedge products each multiplied by positive coefficients, then for generic parameters this will be nonzero everywhere except at the horizons and fixed points of the isometry.  However, we did find that for certain choices of the Kerr-NUT-(A)dS parameters, even this sum of squares vanishes at certain sets of points in the spacetime away from the stationary horizons and fixed points of the isometries.  We believe this vanishing anywhere could be eliminated for any set of the three parameters by choosing a sum of six squared norms of the wedge products of the six pairs of a suitable set of four gradients of scalar polynomial curvature invariants.

One general procedure for calculating a scalar polynomial curvature invariant $I_{(p)}$, vanishing on a stationary horizon, that should generically be positive wherever the Killing vectors span the maximal dimension $m$ of the local isometry group that includes a timelike direction (so that one is not at a stationary horizon or fixed point of the isometry, and so that the $n$ independent directions of the local cohomogeneity are all spacelike) would be to take a set of $n+p$ scalar polynomial curvature invariants, $\{S^{(i)}\}$ for $i$ running from 1 to $n+p$ with sufficiently large $p$.  (For example, $p$ could be the number of parameters for a certain class of spacetimes, such as $p = D-1$ for the Kerr-NUT-(A)dS family of metrics \cite{Chen:2006xh} of even dimension $D$, or $p = D-2$ for odd $D$, for which there are $n+p$ algebraically independent scalar polynomial curvature invariants.)  Then there are $(n+p)!/(n!p!)$ ways of choosing $n$ gradients to form the squared norm of wedge product $n$-form, and one can multiply each by the product of the squared norms of the remaining gradients and then sum over the $(n+p)!/(n!p!)$ choices to get an invariant $I_{(p)}$ that will generically be positive where all the gradients are spacelike, as they generically would be under the conditions above.  Of course, precisely what scalar polynomial curvature invariant $I_{(p)}$ one gets would depend upon which set of $n+p$ scalar polynomial curvature invariants $\{S^{(i)}\}$ one chooses, and some special choices could still conceivably give $I_{(p)} = 0$ even away from stationary horizons and fixed points of the isometry group.

In conclusion, we have found a way to generalize the results of Abdelqader and Lake \cite{Abdelqader:2014vaa} for the Kerr metric to give ways of constructing scalar polynomial curvature invariants that vanish on {\it any} stationary horizon.  These invariants might be useful for numerically estimating the location of horizons once a spacetime settles down to becoming nearly stationary.  They may also be interesting in nonstationary black hole spacetimes for defining `curvature-invariant quasi-horizons' where the squared norm of the appropriate wedge product of gradients of scalar polynomial curvature invariants vanishes, as analogues to apparent horizons, though in the nonstationary case the locations of the `curvature-invariant quasi-horizons' may well depend on what scalar polynomial curvature invariants are chosen for the wedge product of their gradients.

We are grateful to Majd Abdelqader for explaining some aspects of the work of Abdelqader and Lake \cite{Abdelqader:2014vaa} and for confirming that the new syzygy of Eq.\ {\ref{syzygy2}} is indeed a syzygy of the Kerr metric, which we have also now checked along with the others.  We also appreciate helpful advice from Alan Coley, Sigbj{\o}rn Hervik, and Eric Woolgar for revisions of the manuscript.  This research has been supported in part by the Natural Sciences and Engineering Research Council of Canada.


\end{document}